\documentclass[sigconf]{acmart}
\usepackage{enumitem}
\usepackage{footnote}
\usepackage{algorithm, setspace}
\usepackage{tabularx}
\usepackage{multirow}
\usepackage{caption}
\usepackage{subcaption}
\usepackage[noend]{algpseudocode}
\floatname{algorithm}{Algorithm}

\setcopyright{none}
\renewcommand\footnotetextcopyrightpermission[1]{}

\AtBeginDocument{%
  \providecommand\BibTeX{{%
    \normalfont B\kern-0.5em{\scshape i\kern-0.25em b}\kern-0.8em\TeX}}}

\begin{document}

\title{Learned Accelerator Framework for Angular-Distance-Based High-Dimensional DBSCAN}

\author{Yifan Wang}
\affiliation{%
  \institution{University of Florida}
  \city{Gainesville, FL}
  \country{USA}
}
\email{wangyifan@ufl.edu}

\author{Daisy Zhe Wang}
\affiliation{%
  \institution{University of Florida}
  \city{Gainesville, FL}
  \country{USA}
}
\email{daisyw@ufl.edu}

\begin{abstract}
  Density-based clustering is a commonly used tool in data science. 
  Today many data science works are utilizing high-dimensional neural embeddings. However, traditional density-based clustering techniques like DBSCAN have a degraded performance on high-dimensional data.
  In this paper, we propose LAF, a generic learned accelerator framework to speed up the original DBSCAN and the sampling-based variants of DBSCAN on high-dimensional data with angular distance metric. This framework consists of a learned cardinality estimator and a post-processing module. The cardinality estimator can fast predict whether a data point is core or not to skip unnecessary range queries, while the post-processing module detects the false negative predictions and merges the falsely separated clusters. 
  The evaluation shows our LAF-enhanced DBSCAN method outperforms the state-of-the-art efficient DBSCAN variants on both efficiency and quality.  
\end{abstract}

\maketitle

\section{Introduction}
\label{sec:intro}
Today's data science research benefits significantly from neural embeddings that are high-dimensional vectors generated by deep neural models. As a widely applied technique in data science, clustering has been associated with embeddings, e.g., \cite{clustering-learn-passage-embeddings-1, clustering-learn-passage-embeddings-2} learn effective passage embeddings with clustering, \cite{lider, metric-space-learned-index} utilize clustering to accelerate the similarity search over embeddings, etc.

As a representative clustering algorithm, Density-Based Spatial Clustering of Applications with Noise (DBSCAN) \cite{dbscan-1996} has a long history of being applied on low-dimensional spatial data (2D or 3D). 
However, DBSCAN usually has a low efficiency, caused by the compute-intensive range queries and becoming more significant on high-dimensional data due to the curse of dimensionality. 
Specifically, DBSCAN considers clusters to be high-density areas separated by low-density areas. Based on this, the algorithm repeatedly expands each cluster to its neighboring high-density areas (where the points are called \textit{core points}) until the cluster is completely surrounded by low-density areas (where each point is either a \textit{non-core point} or a noise). For each point, DBSCAN has to do a heavy range search to determine whether it is core or not, which requires intensive computation and limits its application in large-scale high-dimensional data analysis.
To improve the efficiency of DBSCAN, previous works propose many variants, e.g., sampling-based DBSCAN variants \cite{sampling-based-dbscan-rough-dbscan-viswanath2009, sampling-based-dbscan-dbscanpp, sampling-based-dbscan-sng-dbscan-jiang2020faster, sampling-based-dbscan-luchi2019, sampling-based-BIRCHSCAN-dbscan-Ventorim2021AS} improve the efficiency by executing the heaviest computation within a small subset instead of the whole dataset, some other works reduce the latency by accelerating the range queries in DBSCAN \cite{range-search-speedup-dbscan-lv2016efficient, range-search-speedup-dbscan-weng2021h}, \cite{pruning-dist-dbscan-cheng2021fast, pruning-dist-dbscan-KNN-BLOCK-DBSCAN-chen2019knn, pruning-dist-dbscan-BLOCK-DBSCAN, pruning-dist-dbscan-NQ-DBSCAN-chen2018fast, pruning-dist-dbscan-rho-approx, pruning-dist-dbscan-rho-approx-2} prune unnecessary distance computation during the clustering, etc. But many of them are designed for low to middle dimensional data (mostly less than 100-dimensions). 
Therefore they are still not suitable for high-dimensional neural embeddings with hundreds to thousands of dimensions (e.g., the 768-dim BERT \cite{bert} embeddings).

In this paper, we solve this problem by skipping the range queries for non-core and noise points, given that only the number of neighbors is needed to confirm the point is non-core or noise. And this can be done by cardinality estimation, i.e., the techniques to estimate the number of results before executing a query. 
In multi-dimensional data management, cardinality estimation is usually used to predict the number of neighbors from a distance-based range query without executing it, and the estimation results can help optimize the critical operations like range search and similarity join. 
Traditional cardinality estimation for range queries relies on sampling or kernel density estimation \cite{tutorial-card-est-high-dim}. 
Recent works apply advanced machine learning to it and propose the \textit{learned cardinality estimation} techniques. They are normally based on regression models from non-deep regressors (e.g., XGBoost) to deep regressors (e.g., deep neural networks), whose input is the query point and the distance threshold (i.e., the range), and output is the estimated number of neighbors in that range. By learning the data distribution, learned cardinality estimation makes more accurate prediction than the traditional approaches. 
The state-of-the-art learned cardinality estimation methods deploy various deep regression models, e.g., Convolutional neural network \cite{card-est-sun2021learned}, Recursive Model Index \cite{rmi-kraska2018case, cardnet-wang2020monotonic}, Deep Lattice Network \cite{Deep-lattice-networks-you2017deep}, CardNet \cite{cardnet-wang2020monotonic}, SelNet \cite{selnet-wang2021consistent}, etc., which perform effectively on high-dimensional data. 
And they can achieve high predicting efficiency both theoretically and practically. In theory, when a model structure is fixed, its prediction time complexity is constant with the data scale, while in practice, the models can be significantly accelerated by GPU. The training time is not an issue due to the generalization capability of the neural models, i.e., a trained estimator can be used on any other dataset with similar distribution. 
With a learned cardinality estimator, whether a point is core or not can be predicted before executing the range query, by which the unnecessary computation on non-core and noise points will be effectively reduced.

 Particularly, our approach is designed for clustering based on angular distance, like cosine distance, and in this paper we will not investigate other distance metrics like Euclidean distance, due to two reasons:  
(1) Angular distance is worth being specifically studied. In neural embedding based applications, cosine and Euclidean distances are the dominant distance metrics for measuring embedding similarity. So focusing on cosine distance is enough to benefit a wide range of applications.
(2) Our idea is theoretically more suitable for this metric. Angular distance is usually bounded, e.g., cosine distance is within the range 0 $\sim$ 2, which makes training of the cardinality estimator more effective than using Euclidean distance. 
Specifically, a regressor normally makes better predictions when the training set covers more possible input values, which is hard for Euclidean distance whose value range is infinite (i.e., $0 \sim +\infty$), but easier for the bounded cosine distance. For example, in our evaluation we construct the training set using cosine distance thresholds from 0.1 to 0.9, which is enough to cover most cases.     
Therefore, other distances are out of scope for this paper. However, our method does not have a hard constraint on the distance metric, so we may explore Euclidean distance in future work.

We propose \textbf{LAF}, a generic \textbf{L}earned  \textbf{A}ccelerator  \textbf{F}ramework to speed up DBSCAN and its sampling-based variants based on angular distance. 
LAF enhances the algorithm efficiency by placing an extra cardinality estimation step before each range query. If a point is predicted as non-core/noise, the range query for it will be skipped to reduce the computation. This approach works not only on DBSCAN, but also on its sampling-based variants, as the same kind of computation waste also exists in processing the sampled subset. 
LAF also includes a post-processing module to compensate for the clustering quality loss by detecting and merging the wrongly separated clusters caused by the false negative predictions (i.e., predicting core points as non-core/noise). 
To our best knowledge, we are one of the first studies that improve the efficiency of high-dimensional DBSCAN-like clustering by cardinality estimation. And we have open-sourced the code at \url{https://github.com/wyfunique/LAF-DBSCAN}.

Our LAF-enhanced DBSCAN outperforms the state-of-the-art approximate DBSCAN variants in the evaluation. 
Specifically, LAF-enhanced DBSCAN achieves up to 2.9x speedup for DBSCAN and is 60\% $\sim$ 140\% faster than the state-of-the-art approximate DBSCAN variants, with high clustering quality on high-dimensional vectors, and the selected sampling-based DBSCAN variant is also accelerated significantly by LAF (i.e., up to 6.7x speedup) with only tiny or no quality loss. 
The main contributions of this paper are as follows:
\begin{enumerate}[leftmargin=2em]
    \item We develop LAF, a generic learned accelerator framework to accelerate a wide range of DBSCAN-like clustering algorithms.
    \item We propose a novel efficient high-dimensional DBSCAN algorithm using the framework.
    \item We conduct experiments on popular high-dimensional datasets and show the high performance of our proposed algorithm and the usefulness of LAF.
\end{enumerate}

\section{The approach}
\label{sec:approach}
\subsection{DBSCAN, LAF and enhanced DBSCAN}
Algorithm~\ref{alg:laf-dbscan} shows the DBSCAN pseudocode in black text (the red text is inserted by LAF, which is introduced below). DBSCAN classifies the data points as core, non-core and noise points, depending on the number of their neighbors within a range. Given a distance function $d(\cdot, \cdot)$ and a distance threshold $\epsilon$, DBSCAN does a range query for each data point P to find its neighbors $\mathcal{N} = \{Q | d(P,Q) < \epsilon\}$. If it has at least $\tau$ neighbors, point P is a core point and the current cluster will be expanded to its neighbors; otherwise P is non-core or noise, and the current cluster will not grow from P to its neighbors. When the current cluster cannot grow any more, the next cluster will start from some other un-clustered core points. Such a process is repeated until there are no more unclassified (i.e., \textit{undefined} in Algorithm~\ref{alg:laf-dbscan}) points. The difference between non-core and noise points is whether it has a core point neighbor. If a point itself is not core but has at least one neighbor that is core point, then it is a non-core point and will be part of the cluster boundary, while a noise point has no core point neighbor and will not be classified into any cluster. For simplification, in the rest of this paper we denote both of the non-core and noise points as \textit{stop points} when it is unnecessary to distinguish between them.

\label{sec:laf-and-enhanced-dbscan}
LAF works as a plugin to the target algorithm: (1) The cardinality estimator is placed right before each range query, 
and the range query will be executed only if the current point is predicted as core. 
(2) The post-processing module is inserted at the end of the clustering
to detect false negative predictions which wrongly estimate core points as not, and merge the clusters separated by such false stop points. This is to compensate for the effectiveness loss caused by the prediction error. 
Based on LAF, we implement an efficient high-dimensional DBSCAN, called \textit{LAF-enhanced DBSCAN} (a.k.a, LAF-DBSCAN). We show its pseudocode in Algorithm~\ref{alg:laf-dbscan} and use red text to highlight the inserted lines by LAF, while the other lines are the same as original DBSCAN. We also develop \textit{LAF-DBSCAN++}, an enhanced version of DBSCAN++ \cite{sampling-based-dbscan-dbscanpp} by LAF, to present the capability of LAF for accelerating variants of DBSCAN. The details are discussed in Section \ref{sec:exps}.

Basically, LAF inserts three critical functions: $\operatorname{CardEst}$, UpdatePartialNeighbors and $\operatorname{PostProcessing}$, as well as a map $\mathcal{E}$ recording all the points which are predicted as stop points (called \textit{predicted stop points}) and their \textit{partial neighbors}. Here we use the term ``partial neighbors'' because for each predicted stop point, $\mathcal{E}$ does not record all its neighbors, but only a subset generated by $\operatorname{UpdatePartialNeighbors}$. $\operatorname{CardEst}$ is simply using cardinality estimator to predict number of the range query results, while the other two functions are both for the post-processing.
Note that we do not use the exact value of $\tau$ with $\operatorname{CardEst}$ to predict whether a point is core or not. Instead, $\tau$ is multiplied with a positive factor $\alpha$ to threshold the predicted cardinality, as shown in line 6 and 22 of Algorithm~\ref{alg:laf-dbscan}. Here $\alpha$ is used to adjust the false positive and false negative rates, such that users can control the prediction error and manipulate the speed-quality trade-off. Specifically, when $\alpha$ increases, false negative rate increases as more predictions become lower than the threshold, resulting in higher speed and lower quality. When $\alpha$ decreases, false positive rate increases, leading to lower speed and higher quality.     

If a point is predicted to be stop point (line 6-9, 26-27), the corresponding entry will be added into $\mathcal{E}$, otherwise the range query will be executed and the point will be double checked with the query results. 
At the same time, $\mathcal{E}$ will be updated by $\operatorname{UpdatePartialNeighbors}$ using the query results. Finally the post-processing uses $\mathcal{E}$ to update the clustering results $\mathcal{C}$. 

\setlength{\intextsep}{5pt} 

\begin{algorithm}
\setstretch{0.8}
\small
\caption{LAF-enhanced DBSCAN (LAF-DBSCAN)}
\label{alg:laf-dbscan}
\begin{algorithmic}[1]
\Require{Dataset $\mathcal{D}$, distance function $d(\cdot, \cdot)$, distance threshold $\epsilon$, minimum number of neighbors $\tau$, error factor $\alpha$} 
\Ensure{the map from points to their cluster IDs $\mathcal{C}$}
\State Cluster ID c := 0
\color{red}
\State Map from predicted stop points to partial neighbors $\mathcal{E}$:=$\emptyset$
\color{black}
\For{\textbf{each} point P in $\mathcal{D}$} 
    $\mathcal{C}$(P) := \textit{undefined}
\EndFor
\For{\textbf{each} point P in $\mathcal{D}$} 
    \If{$\mathcal{C}$(P) $\neq$ \textit{undefined}}
        \textbf{continue}
    \EndIf
    \color{red}
    \If{CardEst(P) < $\alpha\tau$}
        \State $\mathcal{C}$(P) := \textit{noise}   
        \If{P not in $\mathcal{E}$} $\mathcal{E}$(P) := $\emptyset$ \EndIf
        \State \textbf{continue}
    \EndIf
    \color{black}
    \State Neighbors $\mathcal{N}$ := RangeQuery($\mathcal{D}$, $d$, P, $\epsilon$)
    \color{red}
    \State $\mathcal{E}$ := UpdatePartialNeighbors(P, $\mathcal{N}$, $\mathcal{E}$)
    \color{black}
    \If{|$\mathcal{N}$| < $\tau$}
        \State $\mathcal{C}$(P) := \textit{noise}
        \State \textbf{continue}
    \EndIf
    \State c := c + 1
    \State $\mathcal{C}$(P) := c
    \State $\mathcal{S}$ := $\mathcal{N}$ - \{P\}
    \For{\textbf{each} point Q in $\mathcal{S}$ }
        \If{$\mathcal{C}$(Q) = \textit{noise}} $\mathcal{C}$(Q) := c \EndIf
        \If{$\mathcal{C}$(Q) $\neq$ \textit{undefined}} \textbf{continue} \EndIf
        \State $\mathcal{C}$(Q) := c
        \color{red}
        \If{CardEst(Q) $\geq \alpha\tau$} 
        \color{black}
            \State $\mathcal{N}$ := RangeQuery($\mathcal{D}$, $d$, Q, $\epsilon$)
            \color{red}
            \State $\mathcal{E}$ := UpdatePartialNeighbors(Q, $\mathcal{N}$, $\mathcal{E}$)
            \color{black}
            \If{|$\mathcal{N}$| $\geq \tau$} $\mathcal{S}$ := $\mathcal{S} \cup \mathcal{N}$ \EndIf
        \color{red}
        \Else 
            \If{Q not in $\mathcal{E}$} $\mathcal{E}$(Q) := $\emptyset$ \EndIf
        \color{black}
        \EndIf
    \EndFor
\EndFor
\color{red}
\State $\mathcal{C}$ := PostProcessing($\mathcal{C}$, $\mathcal{E}$, $\tau$)
\color{black}
\State \textbf{return} $\mathcal{C}$
\end{algorithmic}
\end{algorithm}

\setlength{\intextsep}{4pt} 

\begin{algorithm}
\setstretch{0.8}
\small
\caption{UpdatePartialNeighbors}
\label{alg:UpdatePartialNeighbors}
\begin{algorithmic}[1]
\Require{Data point P, its neighbors $\mathcal{N}$, the map $\mathcal{E}$} 
\Ensure{the updated $\mathcal{E}$}
\For{\textbf{each} neighbor P$_n$ in $\mathcal{N}$} 
    \If{P$_n$ is in $\mathcal{E}$}
        $\mathcal{E}$(P$_n$) := $\mathcal{E}$(P$_n$) $\cup$ \{P\}
    \EndIf
\EndFor
\State \textbf{return} $\mathcal{E}$
\end{algorithmic}
\end{algorithm}

\setlength{\floatsep}{0.3cm}

\begin{algorithm}
\setstretch{0.7}
\small
\caption{PostProcessing}
\label{alg:PostProcessing}
\begin{algorithmic}[1]
\Require{the map $\mathcal{C}$ from point to cluster, the map $\mathcal{E}$, $\tau$} 
\Ensure{the updated $\mathcal{C}$}
\For{\textbf{each} point P in $\mathcal{E}$} 
    \If{|$\mathcal{E}$(P)| $\geq \tau$}
        \State Randomly select a non-noise neighbor P$^\prime$ in set $\mathcal{E}$(P) 
        \State Destination cluster ID $c^\prime$ := $\mathcal{C}$(P$^\prime$)
        \State Merge the clusters of $\mathcal{E}$(P) into the destination cluster.
    \EndIf
\EndFor
\State \textbf{return} $\mathcal{C}$
\end{algorithmic}
\end{algorithm}

\setlength{\textfloatsep}{0.1cm}

\subsection{Post-processing strategy}
\label{sec:post-proc}
$\mathcal{E}$ records the \textit{partial neighbors} (i.e., a subset of the true neighbors) for each predicted stop point.
It is filled by $\operatorname{UpdatePartialNeighbors}$ in such a way (as shown in Algorithm~\ref{alg:UpdatePartialNeighbors}): if a predicted stop point P$_n$ is found by another point P as neighbor, then P is also neighbor of P$_n$ and will be added to $\mathcal{E}$(P$_n$). 
Function $\operatorname{PostProcessing}$ (Algorithm~\ref{alg:PostProcessing}) detects the false predicted stop points and merges the clusters separated by those points. Specifically, a point P in $\mathcal{E}$ is a false negative if it has at least $\tau$ partial neighbors (line 2). In such case $\operatorname{PostProcessing}$ will randomly select a 
cluster around it
as the destination cluster (line 3-4), and merge the rest wrongly separated clusters to the destination (line 5).

\section{Experiments}
\label{sec:exps}
\subsection{Experiment settings}
\label{sec:exp-settings}
\textbf{Environment:} A Lambda Quad workstation with 28 3.30GHz Intel Core i9-9940X CPUs, 4 RTX 2080 Ti GPUs and 128 GB RAM. 

\noindent\textbf{Datasets: }
Table \ref{tab:datasets} provides an overview for our evaluation datasets, reporting their sizes, data dimensions, error factors used in evaluation, and their vector types. We introduce more details here: 
\begin{enumerate}
    \item NYTimes:
    300k bag-of-words vectors of NYTimes news articles. We randomly sample 150k vectors from them, normalize the samples and reduce their dimension to 256 through Gaussian random projection, which is the same way as \textit{ANN-benchmark}\protect\footnote{ \protect\url{https://github.com/erikbern/ann-benchmarks}}. The resulting dataset is named \textit{NYT-150k}.
    \item Glove:
    1.2M word embeddings (200-dimensional) pre-trained on tweets. We sample 150k vectors from them and name the sampled dataset \textit{Glove-150k}.
    \item MS MARCO \cite{msmarco-nguyen2016ms}: a benchmarking dataset for passage retrieval, including 8.8M passages. 
    We follow a similar way to \cite{lider} to process this dataset, i.e., generating a 768-dimensional embedding for each passage using the same deep model as \cite{lider}, sampling the embeddings into several subsets and naming them as ``MS-'' followed by the size (e.g., ``MS-100k'' includes around 100k embeddings). In this paper we sample 3 datasets, MS-50k, MS-100k and MS-150k.
\end{enumerate}
In addition, we normalize all the data vectors and split each dataset into training and testing sets by a ratio of 8:2.
For each dataset, we first train the learned cardinality estimator on the training set, then evaluate all the methods on the corresponding testing set, i.e., all the reported experiment results are collected on those testing sets.

\begin{table}[h]
\small
\centering
\begin{tabularx}{\columnwidth}{l|l|l|l|l}
\toprule
 \textbf{Dataset} & \textbf{\#Points} & \textbf{Dim} & \textbf{$\alpha$} & \textbf{Type} \\ 
\toprule
NYT-150k  & 150,000 & 256 & 1.15 & 	Bag-of-words \\
Glove-150k   & 	150,000 & 200 & 2.0 & 	Word embedding\\
MS-150k   & 	152,185  & 768 & 7.7 & 	Passage embedding\\
MS-100k  & 	107,400  & 768 & 2.0 & 	Passage embedding\\
MS-50k  & 	53,700 & 768 & 1.5 & 	Passage embedding\\

\bottomrule
\end{tabularx}
\caption{Evaluation dataset information, including the number of points (\textit{\#Points}), data dimension (\textit{Dim}), error factor $\alpha$ of LAF-DBSCAN on each of them, and the vector type (\textit{Type}).} 
\vspace{-2mm}
\label{tab:datasets}
\end{table}

\noindent\textbf{Metrics: } As discussed in Section \ref{sec:intro}, the distance metric in the evaluation is cosine distance. 
For some baselines which support Euclidean distance only, 
since all data points are normalized, we use Equation~\ref{eq:cos_to_euc} to convert cosine distance ($d_{cos}$) into Euclidean distance ($d_{euc}$), such that the distances in our methods are equivalent to those in the baselines. For example, by the equation, when $d_{cos} = 0.5$, the equivalent $d_{euc} = 1.0$, so if we set the distance threshold $\epsilon = 0.5$ in our methods, the threshold in the baselines will be set as 1.0.

\vspace{-3mm}
\begin{align}
\label{eq:cos_to_euc}
    d_{euc}(\vec{u}, \vec{v}) = \sqrt{2d_{cos}(\vec{u}, \vec{v})} \ \ \ (\operatorname{if}\  \|\vec{u}\| = \|\vec{v}\| = 1)
\end{align}

The evaluation metrics include efficiency and effectiveness metrics. For efficiency, the metric is elapsed time of clustering (including the cardinality estimator prediction time and excluding its training time).
 For effectiveness, the metrics are (1) adjusted RAND
index (ARI) \cite{adjusted-RAND-index-hubert1985comparing} and (2) adjusted mutual information score (AMI) \cite{adjusted-mutual-information-vinh2010information}, computed against the ground truth. A higher score means a better clustering quality. Here we use the clustering results of original DBSCAN as ground truth.

\noindent\textbf{Our methods: } In addition to LAF-DBSCAN, we also use LAF to accelerate a sampling-based DBSCAN variant, DBSCAN++ \cite{sampling-based-dbscan-dbscanpp}. The resulting method is named \textit{LAF-DBSCAN++}, whose goal is to present that LAF works not only on DBSCAN but also on its sampling-based variants (as mentioned in Section \ref{sec:intro}). So it just acts as an auxiliary method and our major method is still LAF-DBSCAN in the evaluation.   
In both methods, the cardinality estimator model is an RMI \cite{rmi-kraska2018case} with three stages, respectively including 1, 2, 4 fully-connected neural networks from top to bottom stage. Each neural network has 4 hidden layers whose widths are 512, 512, 256, and 128. Such an estimator has been 
used as a strong baseline in \cite{cardnet-wang2020monotonic}, from where we borrow the code directly. On each training set, the cardinality estimator is trained for 200 epochs with batch size 512.

 Though there are also other learned cardinality estimators, like CardNet \cite{cardnet-wang2020monotonic} and SelNet \cite{selnet-wang2021consistent}, we will not explore which estimator is the best for our methods, as it is out of scope for this paper. Specifically, the goal of this paper is to reveal the potential of such a new idea on speeding up DBSCAN, and in our evaluation the RMI has already performed well enough to demonstrate the potential.

\noindent\textbf{Baselines: } 
The baselines are described as follows:
\begin{enumerate}
    \item DBSCAN: the original DBSCAN. Its clustering results are used as the ground truth for other methods.
    \item DBSCAN++ 
    \protect\footnote{code available at \protect\url{https://github.com/jenniferjang/dbscanpp}} 
    \cite{sampling-based-dbscan-dbscanpp}: an approximate DBSCAN variant that speeds up DBSCAN by sampling the dataset and limiting the heaviest computation within the samples.  Specifically, DBSCAN++ samples a subset of data points, within which the core points are detected w.r.t. the entire dataset. Then the clusters first grow around those core points within the subset, and finally all the unclassified points outside the subset are directly assigned to their closest core points. Our LAF-DBSCAN++ method is built on top of DBSCAN++.   
    \item KNN-BLOCK DBSCAN
    \protect\footnote{code available at \protect\url{https://github.com/XFastDataLab/KNN-BLOCK-DBSCAN}} 
    \cite{pruning-dist-dbscan-KNN-BLOCK-DBSCAN-chen2019knn}: an approximate DBSCAN variant which improves efficiency by pruning unnecessary distance computation with K-nearest neighbor queries. We  denote it as ``KNN-BLOCK'' in the tables and figures.
    \item BLOCK-DBSCAN
    \protect\footnote{code available at \protect\url{https://github.com/XFastDataLab/BLOCK-DBSCAN}} 
    \cite{pruning-dist-dbscan-BLOCK-DBSCAN}: a method similar to KNN-BLOCK DBSCAN, but facilitated by cover tree based range queries.
    \item $\rho$-approximate DBSCAN
    \protect\footnote{code and binary available at \protect\url{https://sites.google.com/view/approxdbscan}, we use the version 2.0} 
    \cite{pruning-dist-dbscan-rho-approx, pruning-dist-dbscan-rho-approx-2}: an approximate DBSCAN variant which accelerates DBSCAN by relaxing the density criteria with an approximation factor $\rho$ ($\rho > 0$). 
\end{enumerate}
Our methods and the baselines are all implemented mainly in C++. 

\noindent\textbf{Parameters: }
The key parameters in all experiments (except the trade-off evaluation) are set as follows, while their settings in the trade-off are introduced separately in Section \ref{sec:exp-tradeoff}. 
(1) Distance threshold  $\epsilon$ and neighbor threshold $\tau$ are set dynamically in different experiments, which will be explicitly stated. (2) For LAF-DBSCAN, the error factor $\alpha$ is set in an ad-hoc manner for different datasets, as reported in Table \ref{tab:datasets}. For LAF-DBSCAN++, its $\alpha$ is fixed to be 1.0.
(3) For DBSCAN++, the sample fraction $p$ is automatically set based on the ratio of predicted core points. Specifically, we first get the ratio of points that are predicted as core by the cardinality estimator (denoted by $R_c$), then $p = \delta + R_c$, where $\delta$ is a user-determined offset ranging from 0.1 to 0.3. In our evaluation, the final $p$ normally ranges within 0.2 $\sim$ 0.6. And $p$ of LAF-DBSCAN++ keeps identical to DBSCAN++.
(4) For KNN-BLOCK DBSCAN, we control two parameters of the k-means tree for KNN search: branching factor (set as 10) and ratio of leaves to check (set as 0.6).  (5) For BLOCK-DBSCAN, we control the basis of the cover tree (set as 2) and the maximum iterations when computing the minimum distance between inner core blocks (i.e., \textit{RNT} in \cite{pruning-dist-dbscan-BLOCK-DBSCAN}, set as 10). (6) For $\rho$-approximate DBSCAN, we set $\rho = 1.0$.

\subsection{Representative ($\epsilon$, $\tau$) and proper $\alpha$ selection}
We select the proper $\epsilon$ and $\tau$ according to the \textit{noise ratio}, i.e., the portion of noise points in each dataset. A proper ($\epsilon$, $\tau$) should lead to (1) a low to middle noise ratio and (2) enough number of clusters, since the clustering makes no sense when too many noises exist or most points are grouped into very few clusters.  
So we do a grid search to select the values of ($\epsilon$, $\tau$) which make (1) noise ratio smaller than 0.6 and (2) the number of clusters large than 20 in most datasets. 
Table~\ref{tab:noise-stat} shows part of the statistics, where each cell includes a pair \textit{(noise ratio, number of clusters)} for the corresponding case. The cells satisfying the conditions are highlighted, for example, (0.55, 5) and (0.6, 5) are both proper ($\epsilon$, $\tau$) since either of them makes at least 2 out of 3 datasets satisfy the conditions; while (0.5, 5) and (0.7, 5) should be avoided. Finally, we choose three ($\epsilon$, $\tau$) values to report throughout this paper: (0.5, 3), (0.55, 5) and (0.6, 5).

\begin{table}[!h]
\small
\centering
\begin{tabularx}{0.8\columnwidth}{l|cccc}
\toprule
 $\boldsymbol{(\epsilon, \tau)}$ & MS-50k & MS-100k & MS-150k  \\ 
\midrule
\textbf{(0.5, 3)} & (0.63, 654) &  \textbf{(0.53, 1071)} & \textbf{ (0.47, 1225)} \\
(0.5, 5) & (0.83, 174) & (0.72, 348) &  (0.64, 380) \\
\textbf{(0.55, 5)} & (0.65, 183) & \textbf{(0.48, 223)} & \textbf{ (0.39, 175)} \\
\textbf{(0.6, 5)} & \textbf{(0.38, 92)} & \textbf{(0.21, 70)} &  \textbf{(0.15, 47)} \\
(0.7, 5) & (0.005, 1) & (0.0007, 1) &  (0.0004, 1) \\
\bottomrule
\end{tabularx}

\caption{Part of the statistics about noise ratio and number of clusters. They are collected by running DBSCAN with different $\epsilon$ and $\tau$ on each dataset. In the table each cell below the dataset name is a pair \textit{(noise ratio, total number of clusters)}, and the proper value pairs are highlighted by bold text.}
\vspace{-3mm}
\label{tab:noise-stat}
\end{table}

\begin{table}
\small
\centering
\begin{tabular}{l|l|l|ccc}
\toprule
  & $\boldsymbol{(\epsilon, \tau)}$ & \textbf{Method} & NYT-150k & Glove-150k & MS-150k \\ 
\toprule

\multirow{15}{*}{ARI} 
& \multirow{5}{*}{(0.5,3)} 
& KNN-BLOCK 	& - & 	0.8597 	& \textbf{0.6004} \\
& & BLOCK-DBSCAN 	& - & 	\textbf{0.8825}	& 0.4953 \\
& & DBSCAN++ &  \textbf{0.7933}  & 	0.8129 	& 0.4218  \\
& & LAF-DBSCAN & 0.7731 & 0.8660 	& 0.4134 \\
& & LAF-DBSCAN++ 	& 0.7321 & 	0.7746 	& 0.4113 \\
\cline{2-6}
& \multirow{5}{*}{(0.55,5)} 
& KNN-BLOCK 	& - & 	0.6942 	& 0.1862 \\
& & BLOCK-DBSCAN 	& - & 	0.8508	& 0.2283 \\
& & DBSCAN++ &  1.0 & 	0.7869 	& 0.1321  \\
& & LAF-DBSCAN & 1.0 & 	\textbf{0.8520}& 	\textbf{0.2309} \\
& & LAF-DBSCAN++ 	& 1.0 & 	0.7444 	& 0.1138 \\
\cline{2-6}
& \multirow{5}{*}{(0.6,5)} 
& KNN-BLOCK 	& - & 	0.2665 	& -0.0444 \\
& & BLOCK-DBSCAN 	& - & 	0.6399	& 0.0046 \\
& & DBSCAN++ &  1.0 & 	0.7801 	& \textbf{0.3687} \\
& & LAF-DBSCAN & 1.0 & 	\textbf{0.8797 }& 0.2643 \\
& & LAF-DBSCAN++ 	& 1.0 & 	0.7653 	& 0.3519 \\

\midrule

\multirow{15}{*}{AMI} 
& \multirow{5}{*}{(0.5,3)} 
& KNN-BLOCK 	& - & 	0.4994 	& \textbf{0.4254} \\
& & BLOCK-DBSCAN 	& - & 	0.6613	& 0.3945 \\
& & DBSCAN++ & 	0.6872 	& 0.6369 & 	0.3965  \\
& & LAF-DBSCAN & 	\textbf{0.7050}& \textbf{0.7558}	& 0.4196 \\
& & LAF-DBSCAN++ 	& 0.6245 & 	0.5947 	& 0.3879 \\
\cline{2-6}
& \multirow{5}{*}{(0.55,5)} 
& KNN-BLOCK 	& - & 	0.3391 	& 0.1738 \\
& & BLOCK-DBSCAN 	& - & 	0.6364	& 0.2626 \\
& & DBSCAN++ & 	1.0 	& 0.6578 & 	0.2288  \\
& & LAF-DBSCAN & 1.0 & 	\textbf{0.7554} 	& \textbf{0.3017} \\
& & LAF-DBSCAN++ 	& 1.0 & 	0.6068 	& 0.2210 \\
\cline{2-6}
& \multirow{5}{*}{(0.6,5)} 
& KNN-BLOCK 	& - & 	0.1427 	& 0.0390 \\
& & BLOCK-DBSCAN 	& - & 	0.4988	& 0.1259 \\
& & DBSCAN++ & 	1.0	& 0.7061 & 	\textbf{0.2836}  \\
& & LAF-DBSCAN & 1.0 	& 	\textbf{0.8167} 	& 0.2763 \\
& & LAF-DBSCAN++ 	& 1.0 & 	0.6822 	& 0.2750 \\

\bottomrule
\end{tabular}
\caption{Clustering quality (AMI and ARI scores) of the approximate methods on the three largest datasets}
\label{tab:quality-major-methods}
\end{table}

\begin{table}[h]
\small
\centering
\begin{tabularx}{0.85\columnwidth}{l|cccc}
\toprule
 $\boldsymbol{(\epsilon, \tau)}$ & MS-50k & MS-100k & MS-150k \\ 
\toprule
(0.5, 3) 
& 864.7s/206.6s  & 	3499.8s/882.7s & 	6931.1s/1669.5s 
 \\
(0.55, 5) 
& 854.7s/180.3s  & 	3367.0s/827.6s & 	6595.1s/1936.6s 
 \\
(0.6, 5) 
& 753.3s/219.9s  & 	2817.9s/1041.1s & 	5385.9s/2539.9s 
\\
\bottomrule
\end{tabularx}

\caption{Clustering time of $\rho$-approximate DBSCAN vs. DBSCAN on different dataset scales. Each cell presents a pair of time, ``$t_1$/$t_2$'', where \textit{$t_1$} is the time of $\rho$-approximate DBSCAN and \textit{$t_2$} is that of DBSCAN given the same ($\epsilon$, $\tau$) and dataset.}
\vspace{-3mm}
\label{tab:eff-exp-rho-approx-dbscan}
\end{table}

\begin{figure*}[h]%
    \centering
    \subfloat[\centering $\epsilon = 0.5$, $\tau = 3$.]{{
        \label{fig:efficiency-exp-eps-0.5_minPts-3}
        \includegraphics[width=0.29\textwidth]{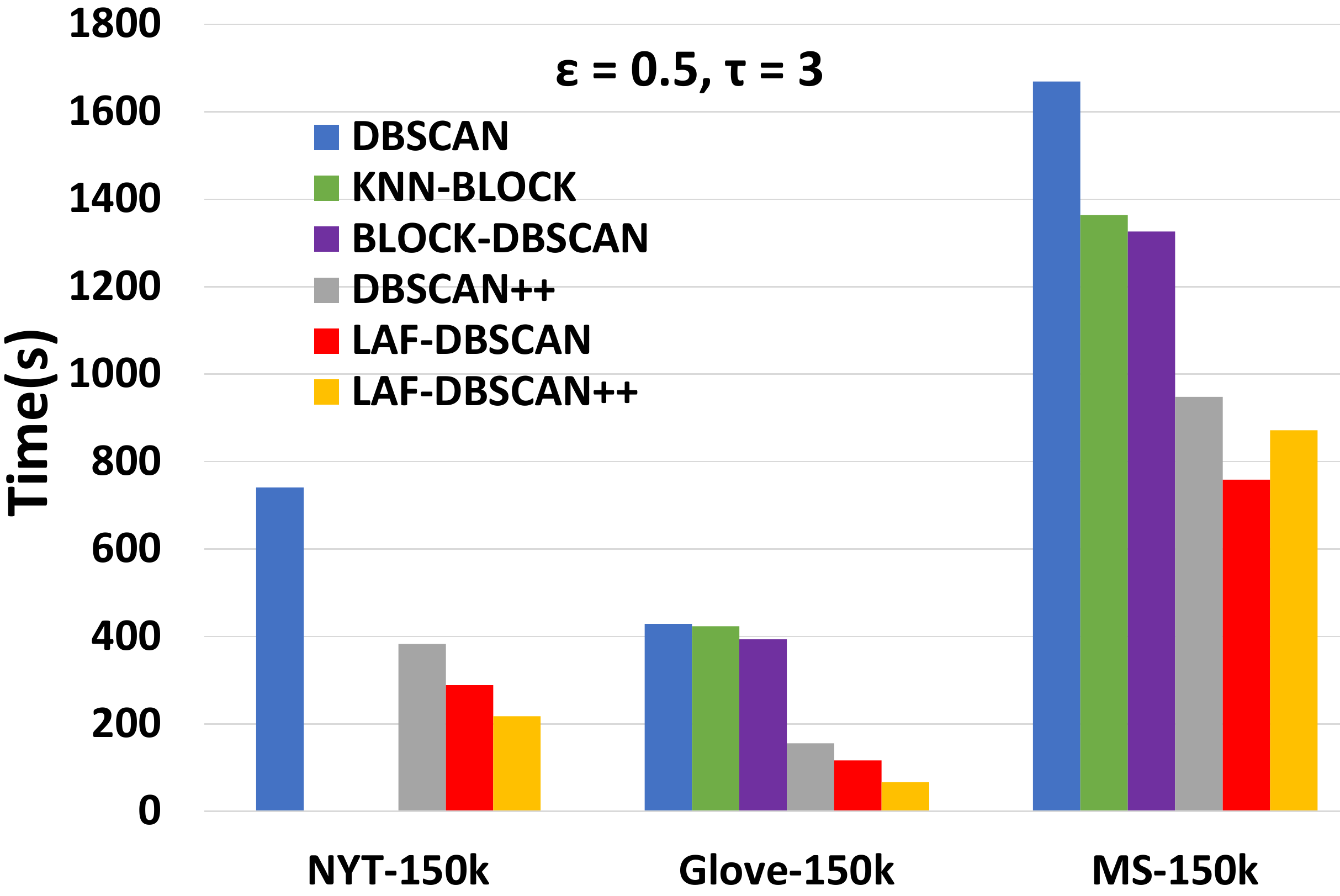}
    }}%
    \qquad
    \subfloat[\centering $\epsilon = 0.55$, $\tau = 5$.]{{
        \label{fig:efficiency-exp-eps-0.55_minPts-5}
       \includegraphics[width=0.29\textwidth]{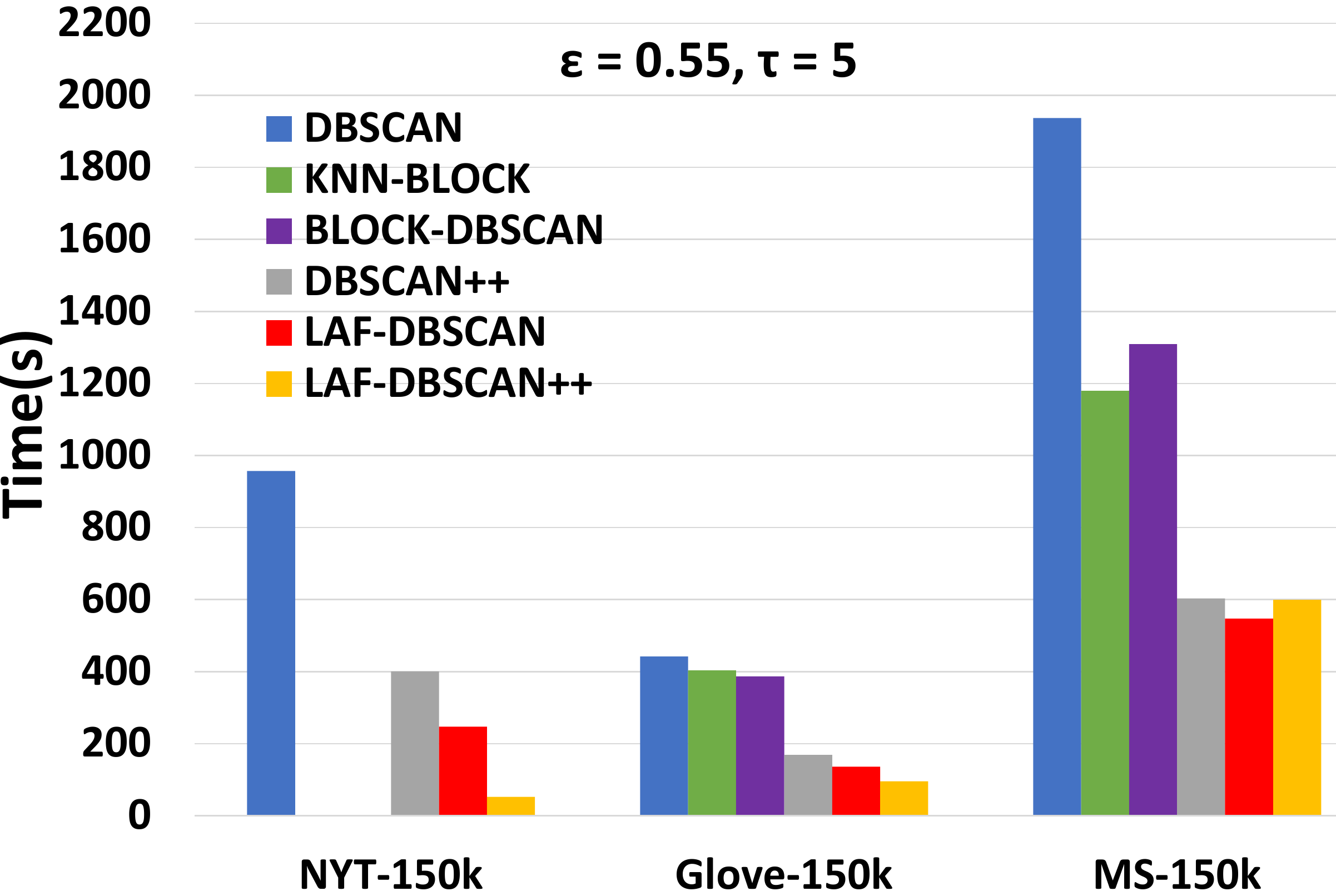}
    }}%
    \qquad
    \subfloat[\centering $\epsilon = 0.6$, $\tau = 5$.]{{
        \label{fig:efficiency-exp-eps-0.6_minPts-5}
        \includegraphics[width=0.29\textwidth]{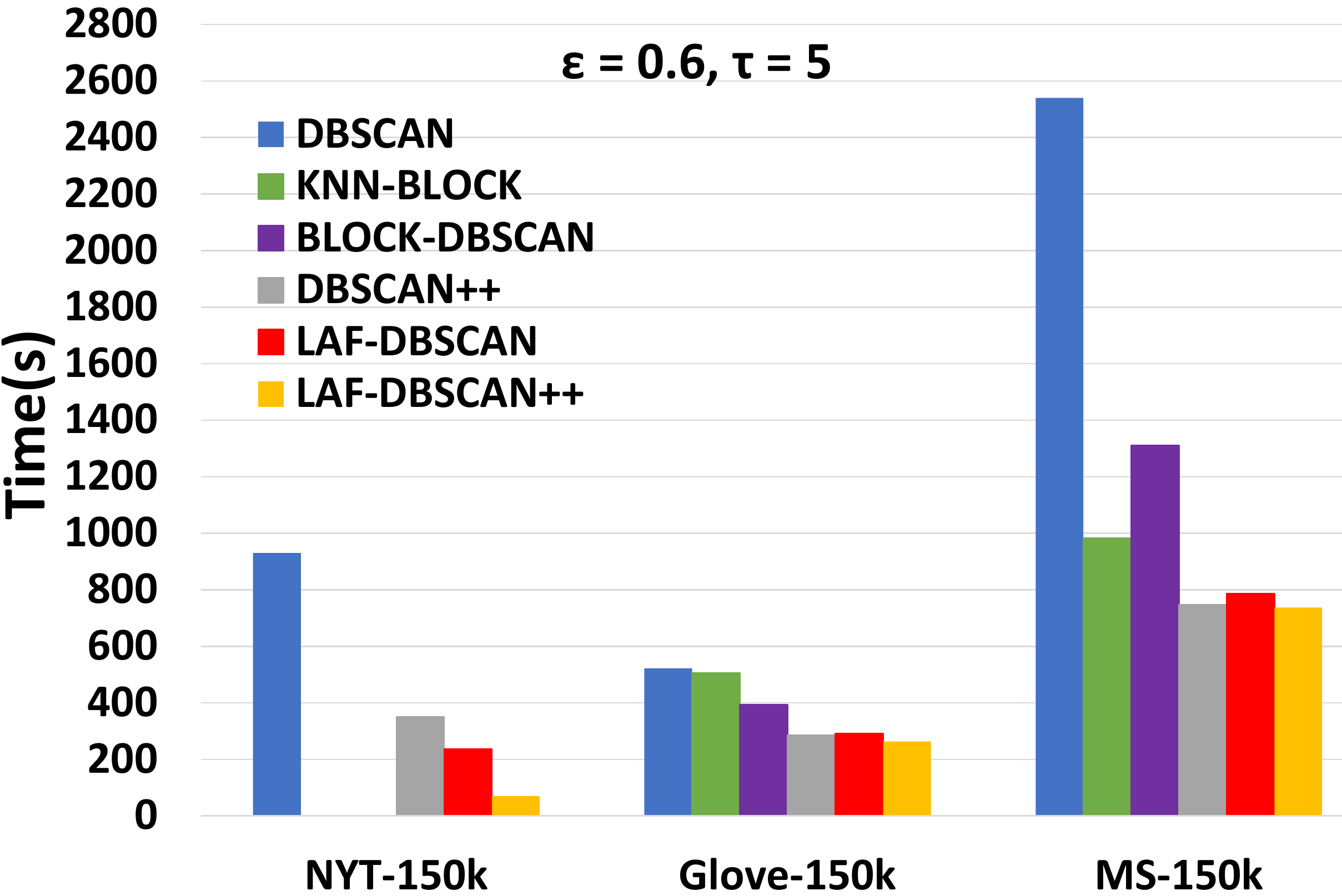}
    }}%
    \caption{Clustering time of all the methods on the three largest datasets}%
    \label{fig:major-eff-exps}%
\end{figure*}

 In this section we also discuss the proper setting of error factor $\alpha$. Basically, there is no quantifiable way to predict the best $\alpha$, as it depends on the dataset. The method we use for this paper is grid search, and our observation can help guide the users: when the vector type is fixed (e.g., dense neural embedding), $\alpha$ should be larger for the larger dataset size or higher data dimension. This can be observed in Table \ref{tab:datasets} on Glove and the three MS datasets. 
The reason is probably the bias in training set. For example, according to Table \ref{tab:noise-stat}, with the increasing data scale, the noise ratio decreases, meaning the fraction of core points increases. Such a bias in training makes the cardinality estimator more likely to predict a larger value. Therefore the $\alpha$ should also increase accordingly.

\subsection{Efficiency and effectiveness evaluation}
\label{sec:effi-and-effe-eval}
We first evaluate the efficiency and effectiveness of each method on the three largest datasets, NYT-150k, Glove-150k and MS-150k.
Table \ref{tab:quality-major-methods} reports the clustering quality via ARI and AMI scores (the higher, the better) for all the approximate methods. As the ground truth, DBSCAN is not included in the table. And Figure \ref{fig:major-eff-exps} illustrates the clustering time of those methods. 
Due to unknown bugs in KNN-BLOCK DBSCAN and BLOCK-DBSCAN, they cannot run on NYT-150k, so their results on NYT-150k are missed in Table \ref{tab:quality-major-methods} and Figure \ref{fig:major-eff-exps}.
Note that we do not include $\rho$-approximate DBSCAN in Table \ref{tab:quality-major-methods}, Figure \ref{fig:major-eff-exps} or any following experiment, due to its significantly low efficiency on high-dimensional data. Specifically, by \cite{pruning-dist-dbscan-rho-approx}, a larger $\rho$ makes $\rho$-approximate DBSCAN more efficient, and $\rho$ ranges from 0.001 to 0.1 in \cite{pruning-dist-dbscan-rho-approx}. However, though we have enlarged $\rho$ to 1.0 
in our evaluation, the method still presents a low efficiency which is even slower than the naive DBSCAN, as shown in Table \ref{tab:eff-exp-rho-approx-dbscan}. This means it suffers much from curse of dimensionality and should not be applied in high-dimensional space. And \cite{pruning-dist-dbscan-BLOCK-DBSCAN} provides further explanation for this problem of $\rho$-approximate DBSCAN.

It is observed that (1) LAF-DBSCAN and LAF-DBSCAN++ achieve the highest efficiency in most cases. For example, LAF-DBSCAN makes up to 2.9x acceleration to DBSCAN as well as reaches 1.6x speed over DBSCAN++, 2.2x speed over KNN-BLOCK DBSCAN and 2.4x speed over BLOCK-DBSCAN. 
(2) LAF-DBSCAN achieves the highest quality in most cases, and in the cases of NYT-150k where the three methods have same scores, LAF-DBSCAN only takes 60\% $\sim$ 70\% time of DBSCAN++, as shown in Figure \ref{fig:major-eff-exps}. (3) LAF-DBSCAN++ usually has a slightly lower quality than DBSCAN++, but gains much more on the efficiency (i.e., up to 6.7x acceleration to DBSCAN++), which makes it practical with better speed-quality trade-off capability than DBSCAN++. 
 (4) Due to curse of dimensionality, all methods perform worse on MS-150k than the other datasets. Specifically, higher dimension usually means more complex distribution. For LAF, the distribution is harder to fit and more false negative predictions (FN) are made, e.g., when $\epsilon=0.5,\tau=3$, the number of FN in NYT/Glove/MS-150k are respectively 5687/2010/7425, which has a negative correlation with the results in Table \ref{tab:quality-major-methods}. More complex distribution also makes sampling less representative and neighbor search less effective, which degrades clustering quality of the baselines too.

\subsection{Speed-quality trade-off evaluation}
\label{sec:exp-tradeoff}
We use MS-150k and Glove-150k with the setting $\epsilon = 0.5, \tau = 3$ to present the speed-quality trade-off capabilities of all the  approximate methods except $\rho$-approximate DBSCAN as discussed in Section \ref{sec:effi-and-effe-eval}.
We adjust the performance of DBSCAN++ and LAF-DBSCAN++ by varying the sample fraction $p$, which is completed by varying the offset $\delta$ (mentioned in Section 3.1) within 0.1 $\sim$ 0.9, while for LAF-DBSCAN the error factor $\alpha$ is varied from 1.1 to 15.0 (which is fixed as 1.0 in LAF-DBSCAN++). For KNN-BLOCK DBSCAN, we vary the branching factor within 3 $\sim$ 20 and the leaves ratio from 0.001 to 0.3. For BLOCK-DBSCAN, we vary the cover tree basis from 1.1 to 5 while fix the maximum iterations as 10.

\begin{figure}[!h]
  \centering
  \includegraphics[width=0.8\columnwidth]{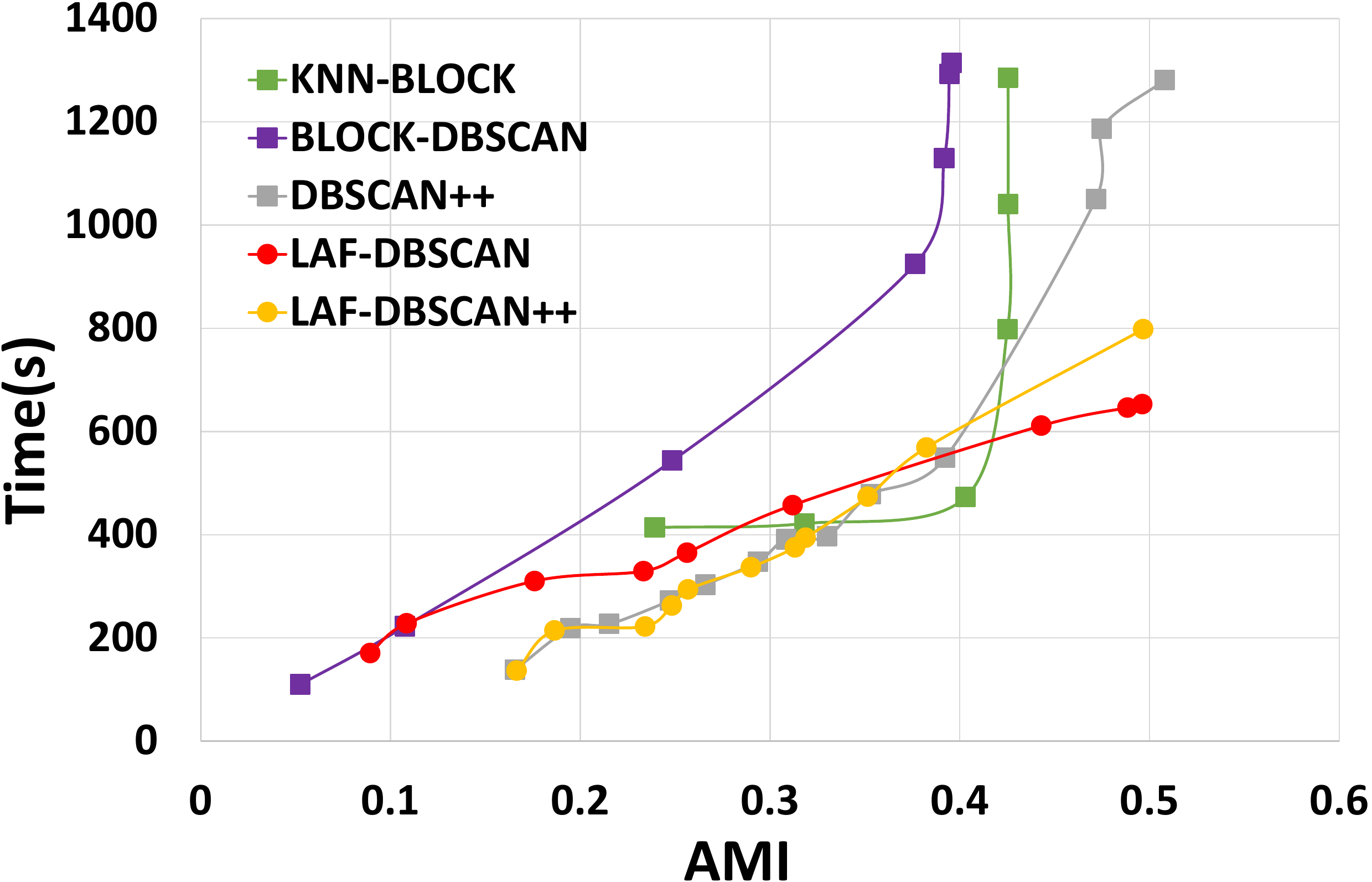}
  \caption{Speed-quality trade-off curves of the approximate methods on dataset MS-150k}
  \setlength{\belowcaptionskip}{-50pt}
  \label{fig:major-tradeoff-ms150k}
\end{figure}

\begin{figure}[!h]
  \centering
  \includegraphics[width=0.8\columnwidth]{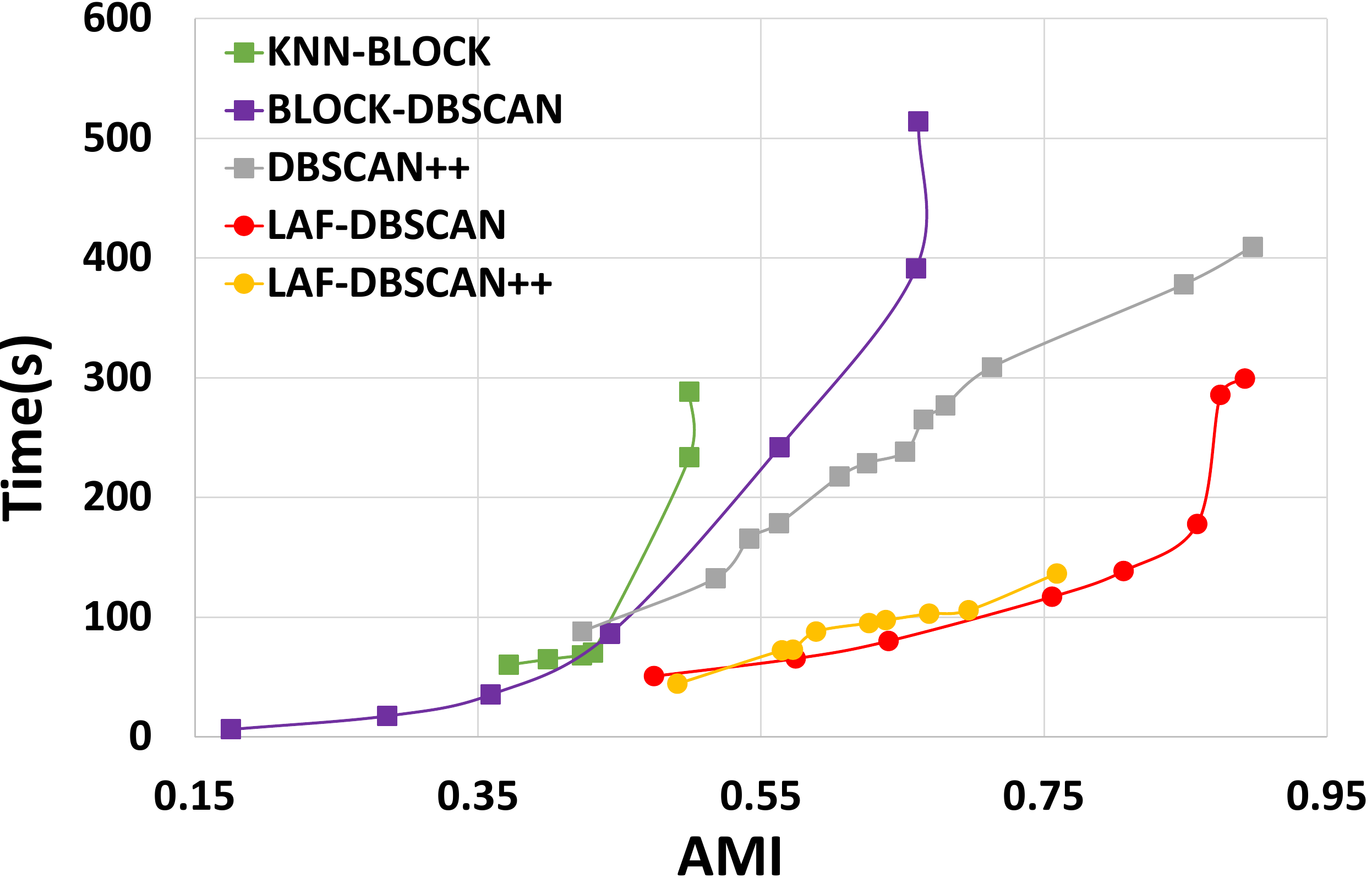}
  \caption{Speed-quality trade-off curves of the approximate methods on dataset Glove-150k}
  \setlength{\belowcaptionskip}{-50pt}
  \label{fig:major-tradeoff-glove}
\end{figure}

As illustrated in Figure \ref{fig:major-tradeoff-ms150k} and \ref{fig:major-tradeoff-glove}, LAF-DBSCAN and LAF-DBSCAN++ have the best speed-quality trade-off capabilities in the high-quality areas (e.g., where AMI > 0.4) on both Glove-150k and MS-150k. Given that the real world normally demands a relatively high clustering quality, we conclude that LAF-DBSCAN and LAF-DBSCAN++ achieve better trade-off capabilities than all the baselines in practice.   
Furthermore, on both datasets, LAF-DBSCAN++ presents a better trade-off than DBSCAN++, meaning that LAF significantly reduces the clustering time of DBSCAN++ with a relatively tiny quality loss, which further proves the strength and usefulness of LAF for a wide range of DBSCAN variants. 

\begin{table}[h]
\small
\centering
\begin{tabularx}{0.9\columnwidth}{l|l|ccc}
\toprule
  & \textbf{Method} & MS-50k & MS-100k & MS-150k \\ 
\toprule

\multirow{5}{*}{\textbf{ARI}} 
& KNN-BLOCK 	& 0.7577 & 	0.3828 	& 0.1862 \\
& BLOCK-DBSCAN 	& \textbf{0.7710} & 	0.4632	& 0.2283 \\
& DBSCAN++ & 0.7238 & 0.4690 & 	0.1321  \\
& LAF-DBSCAN & 0.7581 & \textbf{0.5524} & 	\textbf{0.2309}  \\
& LAF-DBSCAN++ & 0.6455 & 0.3077 & 	0.1138  \\
\midrule

\multirow{5}{*}{\textbf{AMI}} 
& KNN-BLOCK 	& 0.5708 &  0.2736 	& 0.1738 \\
& BLOCK-DBSCAN 	& 0.6134 & 	0.3518	& 0.2626 \\
& DBSCAN++ & 0.6264 & 0.4494 & 	0.2288  \\
& LAF-DBSCAN & \textbf{0.7043} & \textbf{0.5034} & 	\textbf{0.3017}  \\
& LAF-DBSCAN++ & 0.5328 & 0.3197 & 	0.2210  \\

\bottomrule
\end{tabularx}
\caption{Clustering quality of all the approximate methods on datasets of different scales ($\epsilon = 0.55$, $\tau = 5$)}
\label{tab:quality-vs-ds-size}
\vspace{-3mm}
\end{table}

\begin{figure}[!h]
  \centering
  \includegraphics[width=0.8\columnwidth]{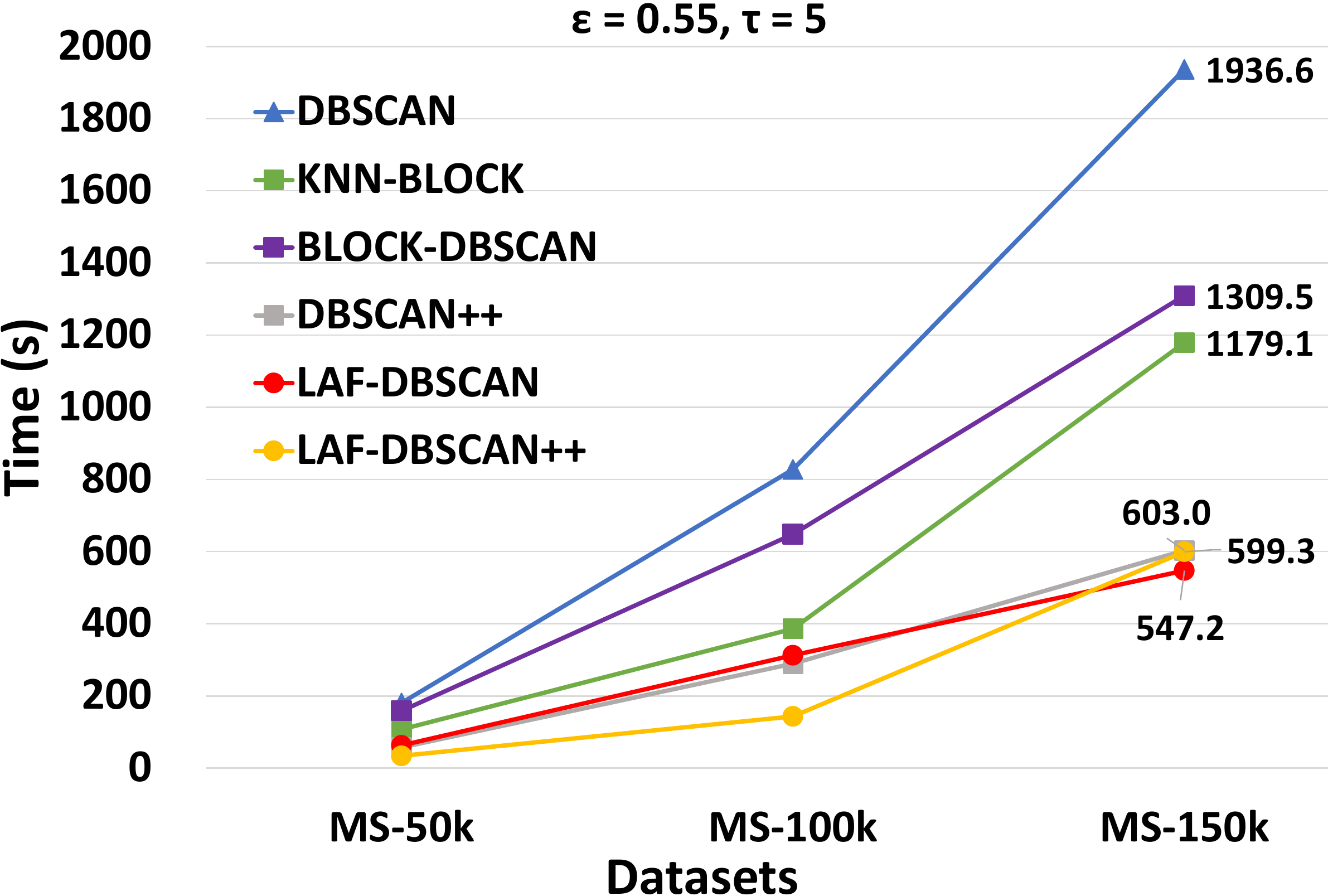}
  \caption{Clustering time of all the methods on datasets of different scales ($\epsilon=0.55$, $\tau=5$)}
  \setlength{\belowcaptionskip}{-50pt}
  \label{fig:eff-exps-zoomed}
\end{figure}

\subsection{Scalability evaluation}
To evaluate the scalability, we run all the methods on the three MS datasets of different scales, and report the results for $\epsilon=0.55$ and $\tau=5$, as the results of other ($\epsilon, \tau$) are similar. 
The quality scores are reported in Table~\ref{tab:quality-vs-ds-size} and the efficiency results are reported in Figure~\ref{fig:eff-exps-zoomed} where we annotate the points of MS-150k with the numbers of the clustering time for a clearer view. They prove that our LAF-enhanced methods are highly effective and scalable, based on these observations: (1) similar to the case of efficiency and effectiveness evaluation, here LAF-DBSCAN still achieves the best quality in most cases, with the highest speed on the largest dataset (whose time is the shortest 547.2s). (2) LAF-DBSCAN has the slowest growth of clustering time when data scale increases, which presents its higher scalability than the baselines. (3) In most cases the quality of LAF-DBSCAN++ is close to DBSCAN++, while in other cases they get closer quickly with the increasing data scale, showing the higher scalability of LAF-DBSCAN++ than DBSCAN++.

\subsection{Missed cluster analysis}
In addition to the wrongly split cluster error discussed in Section \ref{sec:approach}, a cluster may be fully missed if all its core points are falsely predicted to be non-core or noise.  
Fortunately, this fully missed cluster error only has a negligible impact on the quality, as it usually occurs in very tiny clusters.
We choose the cases where LAF-DBSCAN achieves the lowest quality on each dataset according to Table \ref{tab:quality-major-methods} (i.e., ($\epsilon$,$\tau$) = (0.5,3) on NYT-150k, (0.55,5) on Glove-150k and MS-150k) and report the fully missed cluster information in Table \ref{tab:fully-missed-clusters}. Though in the worst cases, LAF-DBSCAN fully misses more than 50\% clusters, it still guarantee the major clusters to be found, since the missed clusters in total include only 1\% $\sim$ 6\% of the non-noise points. And the average size of the missed clusters (ASMC) is too tiny (i.e., only including 3 $\sim$ 7 points on average) to have a non-trivial impact on the overall clustering quality. Therefore, we do not further discuss such an error in this paper.

\begin{table}
\small
\centering
\begin{tabular}{l|l|ccc}
\toprule
 $\boldsymbol{(\epsilon, \tau)}$ & \textbf{Dataset} & MC/TC & MP/TPC & ASMC\\ 
\toprule

(0.5, 3) & NYT-150k & 63/92 & 209/19358	& 3.32  \\
\midrule

(0.55, 5) & Glove-150k & 39/81 & 	250/7879 & 	6.41  \\
\midrule

(0.55, 5) & MS-150k & 159/176 & 1107/18384 & 	6.96  \\
\bottomrule

\end{tabular}
\caption{Statistics for fully missed clusters by LAF-DBSCAN. \textit{MC}, the number of fully \underline{M}issed \underline{C}lusters; \textit{TC}, the \underline{T}otal number of groundtruth \underline{C}lusters; \textit{MP}, the number of \underline{M}issed data \underline{P}oints; \textit{TPC}, the \underline{T}otal number of data \underline{P}oints belonging to the groundtruth \underline{C}lusters, i.e., the non-noise points; \textit{ASMC}, \underline{A}verage \underline{S}ize of the fully \underline{M}issed \underline{C}lusters.}
\vspace{-3mm}
\label{tab:fully-missed-clusters}
\end{table}

\vspace{-3mm}

\section{Conclusion and future work}
To improve efficiency and scalability of high-dimensional DBSCAN-like clustering for angular distance, we propose LAF, a generic learned accelerator framework using learned cardinality estimation techniques to reduce unnecessary range queries in the clustering, and compensating for the quality loss by detecting the false negative and merging the wrongly separated clusters. Our evaluation shows that the LAF-enhanced methods do have a significantly higher efficiency than the state-of-the-art efficient DBSCAN approaches with also high quality, as well as a better speed-quality trade-off capability than the baselines.    
The main limitation of this work is the limited range of applicable distance metrics. 
But since there is no hard constraint on the distance metric, our methods are easy to adapt to other distances, which will be explored in the future work. The future work also includes studying the impact of the cardinality estimator being used, extensively investigating the proper $\alpha$, etc.

\bibliographystyle{ACM-Reference-Format}
\bibliography{main}


\begin{thebibliography}{28}


\ifx \showCODEN    \undefined \def \showCODEN     #1{\unskip}     \fi
\ifx \showDOI      \undefined \def \showDOI       #1{#1}\fi
\ifx \showISBNx    \undefined \def \showISBNx     #1{\unskip}     \fi
\ifx \showISBNxiii \undefined \def \showISBNxiii  #1{\unskip}     \fi
\ifx \showISSN     \undefined \def \showISSN      #1{\unskip}     \fi
\ifx \showLCCN     \undefined \def \showLCCN      #1{\unskip}     \fi
\ifx \shownote     \undefined \def \shownote      #1{#1}          \fi
\ifx \showarticletitle \undefined \def \showarticletitle #1{#1}   \fi
\ifx \showURL      \undefined \def \showURL       {\relax}        \fi
\providecommand\bibfield[2]{#2}
\providecommand\bibinfo[2]{#2}
\providecommand\natexlab[1]{#1}
\providecommand\showeprint[2][]{arXiv:#2}

\bibitem[Chen et~al\mbox{.}(2018)]%
        {pruning-dist-dbscan-NQ-DBSCAN-chen2018fast}
\bibfield{author}{\bibinfo{person}{Yewang Chen}, \bibinfo{person}{Shengyu
  Tang}, \bibinfo{person}{Nizar Bouguila}, \bibinfo{person}{Cheng Wang},
  \bibinfo{person}{Jixiang Du}, {and} \bibinfo{person}{HaiLin Li}.}
  \bibinfo{year}{2018}\natexlab{}.
\newblock \showarticletitle{A fast clustering algorithm based on pruning
  unnecessary distance computations in DBSCAN for high-dimensional data}.
\newblock \bibinfo{journal}{\emph{Pattern Recognition}}  \bibinfo{volume}{83}
  (\bibinfo{year}{2018}), \bibinfo{pages}{375--387}.
\newblock


\bibitem[Chen et~al\mbox{.}(2021)]%
        {pruning-dist-dbscan-BLOCK-DBSCAN}
\bibfield{author}{\bibinfo{person}{Yewang Chen}, \bibinfo{person}{Lida Zhou},
  \bibinfo{person}{Nizar Bouguila}, \bibinfo{person}{Cheng Wang},
  \bibinfo{person}{Yi Chen}, {and} \bibinfo{person}{Jixiang Du}.}
  \bibinfo{year}{2021}\natexlab{}.
\newblock \showarticletitle{BLOCK-DBSCAN: Fast clustering for large scale
  data}.
\newblock \bibinfo{journal}{\emph{Pattern Recognition}}  \bibinfo{volume}{109}
  (\bibinfo{year}{2021}), \bibinfo{pages}{107624}.
\newblock
\showISSN{0031-3203}
\urldef\tempurl%
\url{https://doi.org/10.1016/j.patcog.2020.107624}
\showDOI{\tempurl}


\bibitem[Chen et~al\mbox{.}(2019)]%
        {pruning-dist-dbscan-KNN-BLOCK-DBSCAN-chen2019knn}
\bibfield{author}{\bibinfo{person}{Yewang Chen}, \bibinfo{person}{Lida Zhou},
  \bibinfo{person}{Songwen Pei}, \bibinfo{person}{Zhiwen Yu},
  \bibinfo{person}{Yi Chen}, \bibinfo{person}{Xin Liu},
  \bibinfo{person}{Jixiang Du}, {and} \bibinfo{person}{Naixue Xiong}.}
  \bibinfo{year}{2019}\natexlab{}.
\newblock \showarticletitle{KNN-BLOCK DBSCAN: Fast clustering for large-scale
  data}.
\newblock \bibinfo{journal}{\emph{IEEE transactions on systems, man, and
  cybernetics: systems}} \bibinfo{volume}{51}, \bibinfo{number}{6}
  (\bibinfo{year}{2019}), \bibinfo{pages}{3939--3953}.
\newblock


\bibitem[Cheng et~al\mbox{.}(2021)]%
        {pruning-dist-dbscan-cheng2021fast}
\bibfield{author}{\bibinfo{person}{Difei Cheng}, \bibinfo{person}{Ruihang Xu},
  {and} \bibinfo{person}{Bo Zhang}.} \bibinfo{year}{2021}\natexlab{}.
\newblock \showarticletitle{Fast Density Estimation for Density-based
  Clustering Methods}.
\newblock \bibinfo{journal}{\emph{arXiv preprint arXiv:2109.11383}}
  (\bibinfo{year}{2021}).
\newblock


\bibitem[de~Moura~Ventorim et~al\mbox{.}(2021)]%
        {sampling-based-BIRCHSCAN-dbscan-Ventorim2021AS}
\bibfield{author}{\bibinfo{person}{Igor de Moura~Ventorim},
  \bibinfo{person}{Diego Luchi}, \bibinfo{person}{Alexandre~L. Rodrigues},
  {and} \bibinfo{person}{Fl{\'a}vio~Miguel Varej{\~a}o}.}
  \bibinfo{year}{2021}\natexlab{}.
\newblock \showarticletitle{BIRCHSCAN: A sampling method for applying DBSCAN to
  large datasets}.
\newblock \bibinfo{journal}{\emph{Expert Syst. Appl.}}  \bibinfo{volume}{184}
  (\bibinfo{year}{2021}), \bibinfo{pages}{115518}.
\newblock


\bibitem[Devlin et~al\mbox{.}(2018)]%
        {bert}
\bibfield{author}{\bibinfo{person}{Jacob Devlin}, \bibinfo{person}{Ming-Wei
  Chang}, \bibinfo{person}{Kenton Lee}, {and} \bibinfo{person}{Kristina
  Toutanova}.} \bibinfo{year}{2018}\natexlab{}.
\newblock \bibinfo{title}{BERT: Pre-training of Deep Bidirectional Transformers
  for Language Understanding}.
\newblock
\newblock
\urldef\tempurl%
\url{https://doi.org/10.48550/ARXIV.1810.04805}
\showDOI{\tempurl}


\bibitem[Ester et~al\mbox{.}(1996)]%
        {dbscan-1996}
\bibfield{author}{\bibinfo{person}{Martin Ester}, \bibinfo{person}{Hans-Peter
  Kriegel}, \bibinfo{person}{J{\"o}rg Sander}, \bibinfo{person}{Xiaowei Xu},
  {et~al\mbox{.}}} \bibinfo{year}{1996}\natexlab{}.
\newblock \showarticletitle{A density-based algorithm for discovering clusters
  in large spatial databases with noise.}. In \bibinfo{booktitle}{\emph{kdd}},
  Vol.~\bibinfo{volume}{96}. \bibinfo{pages}{226--231}.
\newblock


\bibitem[Gan and Tao(2015)]%
        {pruning-dist-dbscan-rho-approx}
\bibfield{author}{\bibinfo{person}{Junhao Gan} {and} \bibinfo{person}{Yufei
  Tao}.} \bibinfo{year}{2015}\natexlab{}.
\newblock \showarticletitle{DBSCAN Revisited: Mis-Claim, Un-Fixability, and
  Approximation}. In \bibinfo{booktitle}{\emph{Proceedings of the 2015 ACM
  SIGMOD International Conference on Management of Data}} (Melbourne, Victoria,
  Australia) \emph{(\bibinfo{series}{SIGMOD '15})}.
  \bibinfo{publisher}{Association for Computing Machinery},
  \bibinfo{address}{New York, NY, USA}, \bibinfo{pages}{519–530}.
\newblock
\showISBNx{9781450327589}
\urldef\tempurl%
\url{https://doi.org/10.1145/2723372.2737792}
\showDOI{\tempurl}


\bibitem[Gan and Tao(2017)]%
        {pruning-dist-dbscan-rho-approx-2}
\bibfield{author}{\bibinfo{person}{Junhao Gan} {and} \bibinfo{person}{Yufei
  Tao}.} \bibinfo{year}{2017}\natexlab{}.
\newblock \showarticletitle{On the Hardness and Approximation of Euclidean
  DBSCAN}.
\newblock \bibinfo{journal}{\emph{ACM Trans. Database Syst.}}
  \bibinfo{volume}{42}, \bibinfo{number}{3}, Article \bibinfo{articleno}{14}
  (\bibinfo{date}{jul} \bibinfo{year}{2017}), \bibinfo{numpages}{45}~pages.
\newblock
\showISSN{0362-5915}
\urldef\tempurl%
\url{https://doi.org/10.1145/3083897}
\showDOI{\tempurl}


\bibitem[Hubert and Arabie(1985)]%
        {adjusted-RAND-index-hubert1985comparing}
\bibfield{author}{\bibinfo{person}{Lawrence Hubert} {and}
  \bibinfo{person}{Phipps Arabie}.} \bibinfo{year}{1985}\natexlab{}.
\newblock \showarticletitle{Comparing partitions}.
\newblock \bibinfo{journal}{\emph{Journal of classification}}
  \bibinfo{volume}{2}, \bibinfo{number}{1} (\bibinfo{year}{1985}),
  \bibinfo{pages}{193--218}.
\newblock


\bibitem[Jang and Jiang(2018)]%
        {sampling-based-dbscan-dbscanpp}
\bibfield{author}{\bibinfo{person}{Jennifer Jang} {and}
  \bibinfo{person}{Heinrich Jiang}.} \bibinfo{year}{2018}\natexlab{}.
\newblock \bibinfo{title}{DBSCAN++: Towards fast and scalable density
  clustering}.
\newblock
\newblock
\urldef\tempurl%
\url{https://doi.org/10.48550/ARXIV.1810.13105}
\showDOI{\tempurl}


\bibitem[Jiang et~al\mbox{.}(2020)]%
        {sampling-based-dbscan-sng-dbscan-jiang2020faster}
\bibfield{author}{\bibinfo{person}{Heinrich Jiang}, \bibinfo{person}{Jennifer
  Jang}, {and} \bibinfo{person}{Jakub Lacki}.} \bibinfo{year}{2020}\natexlab{}.
\newblock \showarticletitle{Faster DBSCAN via subsampled similarity queries}.
\newblock \bibinfo{journal}{\emph{Advances in Neural Information Processing
  Systems}}  \bibinfo{volume}{33} (\bibinfo{year}{2020}),
  \bibinfo{pages}{22407--22419}.
\newblock


\bibitem[Kraska et~al\mbox{.}(2018)]%
        {rmi-kraska2018case}
\bibfield{author}{\bibinfo{person}{Tim Kraska}, \bibinfo{person}{Alex Beutel},
  \bibinfo{person}{Ed~H Chi}, \bibinfo{person}{Jeffrey Dean}, {and}
  \bibinfo{person}{Neoklis Polyzotis}.} \bibinfo{year}{2018}\natexlab{}.
\newblock \showarticletitle{The case for learned index structures}. In
  \bibinfo{booktitle}{\emph{Proceedings of the 2018 international conference on
  management of data}}. \bibinfo{pages}{489--504}.
\newblock


\bibitem[Luchi et~al\mbox{.}(2019)]%
        {sampling-based-dbscan-luchi2019}
\bibfield{author}{\bibinfo{person}{Diego Luchi},
  \bibinfo{person}{Alexandre~Loureiros Rodrigues}, {and}
  \bibinfo{person}{Fl{\'a}vio~Miguel Varej{\~a}o}.}
  \bibinfo{year}{2019}\natexlab{}.
\newblock \showarticletitle{Sampling approaches for applying DBSCAN to large
  datasets}.
\newblock \bibinfo{journal}{\emph{Pattern Recognition Letters}}
  \bibinfo{volume}{117} (\bibinfo{year}{2019}), \bibinfo{pages}{90--96}.
\newblock


\bibitem[Lv et~al\mbox{.}(2016)]%
        {range-search-speedup-dbscan-lv2016efficient}
\bibfield{author}{\bibinfo{person}{Yinghua Lv}, \bibinfo{person}{Tinghuai Ma},
  \bibinfo{person}{Meili Tang}, \bibinfo{person}{Jie Cao},
  \bibinfo{person}{Yuan Tian}, \bibinfo{person}{Abdullah Al-Dhelaan}, {and}
  \bibinfo{person}{Mznah Al-Rodhaan}.} \bibinfo{year}{2016}\natexlab{}.
\newblock \showarticletitle{An efficient and scalable density-based clustering
  algorithm for datasets with complex structures}.
\newblock \bibinfo{journal}{\emph{Neurocomputing}}  \bibinfo{volume}{171}
  (\bibinfo{year}{2016}), \bibinfo{pages}{9--22}.
\newblock


\bibitem[Nguyen et~al\mbox{.}(2016)]%
        {msmarco-nguyen2016ms}
\bibfield{author}{\bibinfo{person}{Tri Nguyen}, \bibinfo{person}{Mir
  Rosenberg}, \bibinfo{person}{Xia Song}, \bibinfo{person}{Jianfeng Gao},
  \bibinfo{person}{Saurabh Tiwary}, \bibinfo{person}{Rangan Majumder}, {and}
  \bibinfo{person}{Li Deng}.} \bibinfo{year}{2016}\natexlab{}.
\newblock \showarticletitle{MS MARCO: A human generated machine reading
  comprehension dataset}. In \bibinfo{booktitle}{\emph{CoCo@ NIPS}}.
\newblock


\bibitem[Qin et~al\mbox{.}(2021)]%
        {tutorial-card-est-high-dim}
\bibfield{author}{\bibinfo{person}{Jianbin Qin}, \bibinfo{person}{Wei Wang},
  \bibinfo{person}{Chuan Xiao}, \bibinfo{person}{Ying Zhang}, {and}
  \bibinfo{person}{Yaoshu Wang}.} \bibinfo{year}{2021}\natexlab{}.
\newblock \showarticletitle{High-Dimensional Similarity Query Processing for
  Data Science}. In \bibinfo{booktitle}{\emph{Proceedings of the 27th ACM
  SIGKDD Conference on Knowledge Discovery and Data Mining}} (Virtual Event,
  Singapore) \emph{(\bibinfo{series}{KDD '21})}.
  \bibinfo{publisher}{Association for Computing Machinery},
  \bibinfo{address}{New York, NY, USA}, \bibinfo{pages}{4062–4063}.
\newblock
\showISBNx{9781450383325}
\urldef\tempurl%
\url{https://doi.org/10.1145/3447548.3470811}
\showDOI{\tempurl}


\bibitem[Sun et~al\mbox{.}(2021)]%
        {card-est-sun2021learned}
\bibfield{author}{\bibinfo{person}{Ji Sun}, \bibinfo{person}{Guoliang Li},
  {and} \bibinfo{person}{Nan Tang}.} \bibinfo{year}{2021}\natexlab{}.
\newblock \showarticletitle{Learned cardinality estimation for similarity
  queries}. In \bibinfo{booktitle}{\emph{Proceedings of the 2021 International
  Conference on Management of Data}}. \bibinfo{pages}{1745--1757}.
\newblock


\bibitem[Tian et~al\mbox{.}(2022)]%
        {metric-space-learned-index}
\bibfield{author}{\bibinfo{person}{Yao Tian}, \bibinfo{person}{Tingyun Yan},
  \bibinfo{person}{Xi Zhao}, \bibinfo{person}{Kai Huang}, {and}
  \bibinfo{person}{Xiaofang Zhou}.} \bibinfo{year}{2022}\natexlab{}.
\newblock \bibinfo{title}{A Learned Index for Exact Similarity Search in Metric
  Spaces}.
\newblock
\newblock
\urldef\tempurl%
\url{https://doi.org/10.48550/ARXIV.2204.10028}
\showDOI{\tempurl}


\bibitem[Vinh et~al\mbox{.}(2010)]%
        {adjusted-mutual-information-vinh2010information}
\bibfield{author}{\bibinfo{person}{Nguyen~Xuan Vinh}, \bibinfo{person}{Julien
  Epps}, {and} \bibinfo{person}{James Bailey}.}
  \bibinfo{year}{2010}\natexlab{}.
\newblock \showarticletitle{Information theoretic measures for clusterings
  comparison: Variants, properties, normalization and correction for chance}.
\newblock \bibinfo{journal}{\emph{The Journal of Machine Learning Research}}
  \bibinfo{volume}{11} (\bibinfo{year}{2010}), \bibinfo{pages}{2837--2854}.
\newblock


\bibitem[Viswanath and Babu(2009)]%
        {sampling-based-dbscan-rough-dbscan-viswanath2009}
\bibfield{author}{\bibinfo{person}{P Viswanath} {and} \bibinfo{person}{V~Suresh
  Babu}.} \bibinfo{year}{2009}\natexlab{}.
\newblock \showarticletitle{Rough-DBSCAN: A fast hybrid density based
  clustering method for large data sets}.
\newblock \bibinfo{journal}{\emph{Pattern Recognition Letters}}
  \bibinfo{volume}{30}, \bibinfo{number}{16} (\bibinfo{year}{2009}),
  \bibinfo{pages}{1477--1488}.
\newblock


\bibitem[Wang et~al\mbox{.}(2021a)]%
        {clustering-learn-passage-embeddings-2}
\bibfield{author}{\bibinfo{person}{Xiao Wang}, \bibinfo{person}{Craig
  Macdonald}, \bibinfo{person}{Nicola Tonellotto}, {and} \bibinfo{person}{Iadh
  Ounis}.} \bibinfo{year}{2021}\natexlab{a}.
\newblock \showarticletitle{Pseudo-relevance feedback for multiple
  representation dense retrieval}. In \bibinfo{booktitle}{\emph{Proceedings of
  the 2021 ACM SIGIR International Conference on Theory of Information
  Retrieval}}. \bibinfo{pages}{297--306}.
\newblock


\bibitem[Wang et~al\mbox{.}(2022)]%
        {lider}
\bibfield{author}{\bibinfo{person}{Yifan Wang}, \bibinfo{person}{Haodi Ma},
  {and} \bibinfo{person}{Daisy~Zhe Wang}.} \bibinfo{year}{2022}\natexlab{}.
\newblock \bibinfo{title}{LIDER: An Efficient High-dimensional Learned Index
  for Large-scale Dense Passage Retrieval}.
\newblock
\newblock
\urldef\tempurl%
\url{https://doi.org/10.48550/ARXIV.2205.00970}
\showDOI{\tempurl}


\bibitem[Wang et~al\mbox{.}(2020)]%
        {cardnet-wang2020monotonic}
\bibfield{author}{\bibinfo{person}{Yaoshu Wang}, \bibinfo{person}{Chuan Xiao},
  \bibinfo{person}{Jianbin Qin}, \bibinfo{person}{Xin Cao},
  \bibinfo{person}{Yifang Sun}, \bibinfo{person}{Wei Wang}, {and}
  \bibinfo{person}{Makoto Onizuka}.} \bibinfo{year}{2020}\natexlab{}.
\newblock \showarticletitle{Monotonic cardinality estimation of similarity
  selection: A deep learning approach}. In
  \bibinfo{booktitle}{\emph{Proceedings of the 2020 ACM SIGMOD International
  Conference on Management of Data}}. \bibinfo{pages}{1197--1212}.
\newblock


\bibitem[Wang et~al\mbox{.}(2021b)]%
        {selnet-wang2021consistent}
\bibfield{author}{\bibinfo{person}{Yaoshu Wang}, \bibinfo{person}{Chuan Xiao},
  \bibinfo{person}{Jianbin Qin}, \bibinfo{person}{Rui Mao},
  \bibinfo{person}{Makoto Onizuka}, \bibinfo{person}{Wei Wang},
  \bibinfo{person}{Rui Zhang}, {and} \bibinfo{person}{Yoshiharu Ishikawa}.}
  \bibinfo{year}{2021}\natexlab{b}.
\newblock \showarticletitle{Consistent and flexible selectivity estimation for
  high-dimensional data}. In \bibinfo{booktitle}{\emph{Proceedings of the 2021
  International Conference on Management of Data}}.
  \bibinfo{pages}{2319--2327}.
\newblock


\bibitem[Weng et~al\mbox{.}(2021)]%
        {range-search-speedup-dbscan-weng2021h}
\bibfield{author}{\bibinfo{person}{Shaoyuan Weng}, \bibinfo{person}{Jin Gou},
  {and} \bibinfo{person}{Zongwen Fan}.} \bibinfo{year}{2021}\natexlab{}.
\newblock \showarticletitle{$ h $-DBSCAN: A simple fast DBSCAN algorithm for
  big data}. In \bibinfo{booktitle}{\emph{Asian Conference on Machine
  Learning}}. PMLR, \bibinfo{pages}{81--96}.
\newblock


\bibitem[You et~al\mbox{.}(2017)]%
        {Deep-lattice-networks-you2017deep}
\bibfield{author}{\bibinfo{person}{Seungil You}, \bibinfo{person}{David Ding},
  \bibinfo{person}{Kevin Canini}, \bibinfo{person}{Jan Pfeifer}, {and}
  \bibinfo{person}{Maya Gupta}.} \bibinfo{year}{2017}\natexlab{}.
\newblock \showarticletitle{Deep lattice networks and partial monotonic
  functions}.
\newblock \bibinfo{journal}{\emph{Advances in neural information processing
  systems}}  \bibinfo{volume}{30} (\bibinfo{year}{2017}).
\newblock


\bibitem[Zhan et~al\mbox{.}(2022)]%
        {clustering-learn-passage-embeddings-1}
\bibfield{author}{\bibinfo{person}{Jingtao Zhan}, \bibinfo{person}{Jiaxin Mao},
  \bibinfo{person}{Yiqun Liu}, \bibinfo{person}{Jiafeng Guo},
  \bibinfo{person}{Min Zhang}, {and} \bibinfo{person}{Shaoping Ma}.}
  \bibinfo{year}{2022}\natexlab{}.
\newblock \showarticletitle{Learning Discrete Representations via Constrained
  Clustering for Effective and Efficient Dense Retrieval}. In
  \bibinfo{booktitle}{\emph{Proceedings of the Fifteenth ACM International
  Conference on Web Search and Data Mining}} (Virtual Event, AZ, USA)
  \emph{(\bibinfo{series}{WSDM '22})}. \bibinfo{publisher}{Association for
  Computing Machinery}, \bibinfo{address}{New York, NY, USA},
  \bibinfo{pages}{1328–1336}.
\newblock
\showISBNx{9781450391320}
\urldef\tempurl%
\url{https://doi.org/10.1145/3488560.3498443}
\showDOI{\tempurl}


\end{thebibliography}

\end{document}